# The reception of relativity in the Netherlands[1]

Jip van Besouw and Jeroen van Dongen

*Institute for History and Foundations of Science and Descartes Centre,*
*Utrecht University, Utrecht, the Netherlands*

## 1. Introduction

Albert Einstein published his definitive version of the general theory of relativity in 1915, in the middle of the First World War. The war greatly impeded the international dispersion of his work, as relations and communications between scientists in Allied and Central Power countries had largely broken down. Nevertheless, Einstein and his theory rose to fame shortly afterwards, when in November of 1919 British astronomers claimed to have observationally verified the gravitational bending of light. Relativity still encountered difficulties in gaining acceptance in many quarters, however. In Germany, populist and academic 'anti-relativist' circles protested Einstein's work. They were triggered by his pacifism, and the fact that he was a democrat and a Jew; Einstein further embodied a new kind of highly mathematized physics that seemed to marginalize the nineteenth century universalist experiment-cum-theory gentleman physicist.[2] In France, too, political and professional resentments coloured the reception of both relativity and Einstein.[3]

In the Netherlands, the country that we focus on in this article, interest in Einstein's relativity picked up considerably at the end of 1912. In Germany relativity was already more or less the leading interpretation of electrodynamics among theorists in 1911, but in the Netherlands H.A. Lorentz' version still dominated at the time. There were also strong local rival theories in Britain and France, and broad professional support for Einstein's theory began considerably later there;[4] at the same time, interest in the theory was accompanied by strong public responses.[5] As we will discuss here, these were much more moderate in Holland. Even Dutch anti-relativists (only two have been identified as such in the literature so far: engineer M.W. Polak and well-known philosopher Gerard Heymans) struck a different tone, as A.J. Kox has pointed out: "they remained polite and were careful to avoid any impression of mounting a personal attack against Einstein and his fellow relativists."[6]

---

[1] Published as: pp. 89-110 in *Physics as a calling, science for society. Studies in honor of A.J. Kox*, A. Maas and H. Schatz (eds.) Leiden: Leiden University Press, 2013.
[2] See Goenner (1993), Rowe (2006), Van Dongen (2007), Wazeck (2009).
[3] Reactions to Einstein's visit to France are discussed in Biezunski (1987) and CPAE 13.
[4] Paty (1987), Warwick (2003).
[5] For France, see Moatti (2007), van Kimmenade (2010); for Britain see e.g. Stanley (2003).

One of the issues that we wish to address in this article is how the Dutch reception history of relativity compares to that in other countries: how does the Dutch case resemble or depart from the familiar stories of the larger European nations? Part of the unique nature of the Dutch reception is of course due to Leiden's physicists, who played an important role in the genesis of both the special and the general theory of relativity. In this article, however, we wish to look at the way broader audiences than academic physics professionals viewed Einstein and his theory. Can we identify the factors that shaped the broader Dutch reception of relativity? How particular to the Dutch case were they? Holland had remained neutral during the First World War, and anti-German sentiments were considerably less strong than elsewhere. Did the dynamics of the World War and its aftermath still play a role in shaping public views on relativity? Of course, the Dutch physics elite was personally close to Einstein. To see how that could influence public perceptions of Einstein, we should first address how they themselves were viewed by the public. What relevant values were shared between university professors and other sectors of society? What was the status of science? In this essay we will take up these issues to see if they can shed light on the Dutch reception case. *Vice versa*, we hope that a closer look at the reception of relativity can also teach us things about the public status of science in the interwar years.

The context for this article is given by three studies that focus on the reception of relativity in the Netherlands. The first, by A.J. Kox,[7] primarily discusses the reception of general relativity among Leiden's physicists, and contains a brief discussion of opposition to relativity. The second, a book and Ph.D. thesis by Henk Klomp, *De relativiteitstheorie in Nederland*, goes into more detail regarding the broader reception of the theory. Klomp relates how relativity influenced debates on the certainty of knowledge and the democratic ordering of society. He particularly hones in on how the theory played a role in discussions on secondary education between, foremost historian of science E.J. Dijksterhuis on the one hand, and educational reformers Philipp Kohnstamm and Tatiana Ehrenfest-Afanassjewa on the other.[8] We will return to Klomp's account later. The third article is by one of us,[9] and touches, among other issues, on the Dutch reception of Einstein and relativity in comparison with German events, with a focus on Einstein's Leiden chair and the immediate period after Eddington's announcement. The current paper can be seen as a short elaboration of this account.

---

[6] Kox (1992), pp. 46-47. Leiden's Hegelian philosopher G.J.P.J. Bolland did inveigh against relativity, as an extension of his attacks on Lorentz and physics in general, but Bolland was a notorious and somewhat indiscriminate troublemaker (Otterspeer, 1995).
[7] Kox (1992).
[8] Klomp (1997).
[9] Van Dongen (2012).

## 2. Einstein's chair and Dutch internationalism

Let us first begin by addressing, very briefly, the reception of relativity and Einstein in academic circles. Einstein, of course, had already become friendly with Lorentz in 1911, when he visited Leiden, and had been exchanging letters with him since 1909. Yet, perhaps surprisingly, their correspondence did not discuss relativity in any substantial way before 1912.[10] Lorentz was a towering presence in the Dutch scientific community and this showed, not only in his position as Chairman of the Sciences Section of the Royal Academy of Sciences in Amsterdam, since 1910, but also in the status of his electromagnetic theory. Lorentz understood at an early stage what the differences were between his work and Einstein's theory, and participated in the co-production of what came to be understood as 'relativity theory' while it took shape in the German literature. Nonetheless, he continued to prefer his own version of electrodynamics.[11] Local scholarship tended to follow him in this preference, as is shown for example in the comparison of Lorentz' theory with Einstein's by J.D. van der Waals jr. in 1909[12], and Lorentz initially remained the primary Dutch authority on relativity.[13]

The notion that Einstein's theory could epistemologically be preferred seems to have come to the Netherlands with the arrival of Paul Ehrenfest. Ehrenfest had been involved in subtle German debates about relativistic rigid bodies, and was acutely aware of the theory's interpretative benefits and complexities, as is exhibited in his impressive inaugural lecture in Leiden, held in December of 1912.[14] Lorentz and Ehrenfest began corresponding with Einstein on proto versions of general relativity,[15] and in the period between 1915 and 1920, Leiden grew into a—if not *the*—early hub for relativity research. A steady stream of publications on the subject by W. de Sitter, J. Droste, A.D. Fokker, J.A. Schouten, H.A. Kramers, G. Nordström and others came from Holland.[16]

In 1920, Einstein himself held an inaugural lecture in Leiden after he had been appointed 'Special Professor' there. His friendships with Lorentz and Ehrenfest were of key importance in eventually accepting a tie with Leiden (he had actually turned down full professorships in

---

[10] The correspondence between Einstein and Lorentz can be found in Kox (2008); it began in 1909 and before 1912 it mostly dealt with problems in radiation theory.
[11] See e.g. Staley (1998), pp. 288-289; on the differences between Einstein and Lorentz, see for example Janssen (2002).
[12] Van der Waals jr. (1909).
[13] See for instance how de Sitter (1911, p. 389) acknowledged discussions with Lorentz.
[14] See Ehrenfest (1912), Klein (1970), pp. 1-5; 151-154.
[15] Kox (1988), CPAE 5, CPAE 8.
[16] See Kox (1992) for an overview.

Leiden twice before[17]), and, likely in deference to Lorentz, Einstein reinstated the spacetime metric as a kind of "Lorentzian ether" in his lecture.[18] Einstein's appointment had been held up considerably, as the Dutch authorities had mistaken him for the art critic and communist Carl Einstein: the delay is to be explained by political fears for communist revolutionaries.[19] Einstein was, of course, proposed as a special professor first and foremost for his physics. Consideration of Einstein's political and personal views and those of his Leiden hosts may however assist us in understanding the broader Dutch reception history.

The two Leiden professors who were most directly involved in securing Einstein's appointment, Lorentz and Professor of International Law Cornelis van Vollenhoven, also were leading figures in the Dutch Academy. They were particularly responsible for its efforts to arrive at a reconciliation between the scientific communities of the formerly warring nations. Upon joining the newly created International Research Council in 1919, the Dutch Academy began to lobby for the admittance of sister academies from former Central Power countries that had been barred from membership. For Lorentz, this was a continuation of the reconciliation efforts he had undertaken during the war. Van Vollenhoven had seen a particular role for the Dutch in establishing international peace as early as 1910 and had expressed this in an influential article in *De Gids*.[20] Both believed that the Netherlands could do its duty and further its own position by acting as a peace broker. They acted on this belief in their stewardship of the Dutch Academy's attempts at diplomacy.[21]

Einstein's pacifism and his ideal of a supranational scientific community thus resonated with these Leiden professors in their self-chosen role of peace brokers.[22] Indeed, just after accepting his special professorship, Einstein participated in an event organized through the Paris based *Institut du Froid* at Kamerlingh Onnes' laboratory in which Frenchmen Pierre Weiss and Paul Langevin also participated.[23] As Ehrenfest expressed it, Einstein's role as special professor in Leiden "will contribute enormously to the reestablishment of many disrupted scientific relations in an undemonstrative, yet therefore all the more powerful way."[24]

---

[17] Kox (1993); Einstein to Paul Ehrenfest, 12 September 1919, Doc. 103, CPAE 9; van Dongen (2012).
[18] Einstein (1920), p. 13.
[19] See van Dongen (2012).
[20] Van Vollenhoven (1910).
[21] On the academic rift at the time of the creation of the IRC, see Kevles (1971); on the mediating role of the Dutch, see Otterspeer & Schuller (1997), van Lunteren (2006), Somsen (2012).
[22] For Einstein's positions, see e.g. Rowe & Schulmann (2007), pp. 85-88; on their relation to those of his Leiden colleagues, see van Dongen (2012).
[23] For the science discussed at the event, see Sauer (2007).
[24] Paul Ehrenfest to Albert Einstein, 9 December 1920, CPAE 10, pp. 286-289.

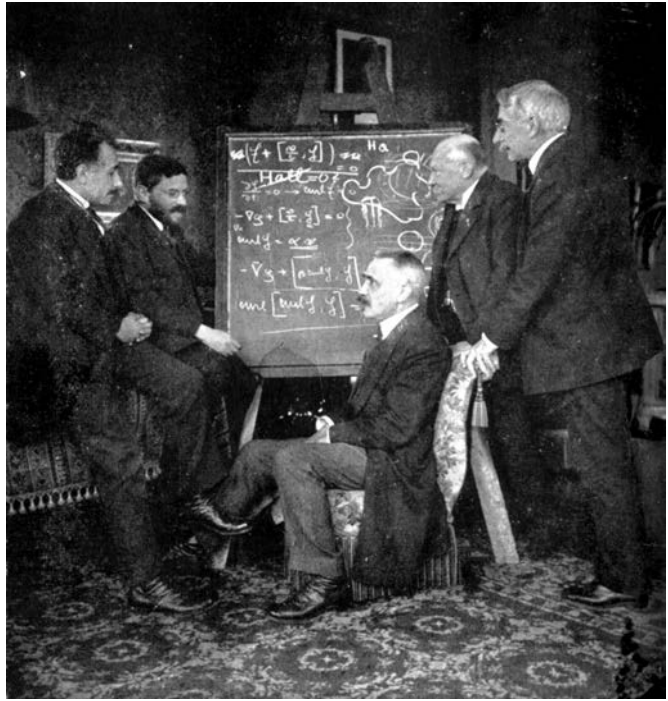

Albert Einstein, Paul Ehrenfest, Paul Langevin, Heike Kamerlingh Onnes and Pierre Weiss discussing problems in condensed matter physics in Onnes' home in Leiden in October of 1920. *Source*: Museum Boerhaave, Leiden.

Dutch neutrality had a long history. It had been inspired by the vulnerable position of the Netherlands between the great European powers, and by a fear for its colonial interests.[25] In the late nineteenth century, Dutch intellectuals had grown afraid of domination by a recently unified Germany: a mediating role between the larger cultures of Europe could ensure Dutch independence and international relevance.[26] Already before the war, publicists and politicians had singled out endeavours for establishing peace and international law as a particularly Dutch moral obligation. The 1899 and 1907 peace conferences in The Hague had been instrumental for this development. Van Vollenhoven, who had been enlisted by the government to participate in organizing a third conference that was supposed to take place in 1915, had played up nationalist sentiments by emphasizing the Dutch heritage in international law that dates back to Hugo Grotius; here, surely, was a task for which the Dutch should unite and a way for the nation to find new prominence.[27] Although the role of neutral countries as moral arbitrators between belligerents had become internationally compromised during the war, public opinion in the Netherlands still favoured policies of neutrality and internationalism.[28] Appointing Einstein would thus assist Lorentz and van Vollenhoven in placing Holland before the eyes of the international academic community as an indispensable peace broker, and Dutch scientists before the eyes of the Dutch public as prominently taking part in the realization of national ambitions.

---

[25] Voorhoeve (1985), pp. 32-35.
[26] Otterspeer & Schuller (1997).
[27] Corduwener (2012).
[28] For a nuanced treatment, see Tames (2006), pp. 262-265.

**3. Science on the defensive**

So, ideals of neutrality and internationalism were shared between scientists and the public alike. Scientists were generally held in high regard, and their cultural status was unchallenged. Professorial appointments were regularly and enthusiastically reported in the press, and summaries of academic lectures, including technical presentations at the Academy, could be found in newspapers like the *Algemeen Handelsblad* or the *Nieuwe Rotterdamsche Courant*. Partly responsible for this state of affairs was, of course, the international prestige of Dutch researchers: between 1901 and 1913, no less than five Dutch scientists had won a Nobel Prize.[29]

The relations between science and society changed near the end of the war. The number of students had steadily increased since the turn of the century. This increase was accelerated by a new law in 1917 that made studies in the sciences even more accessible: proficiency in classical languages and a *gymnasium* diploma were no longer required. As a consequence of the increased enrolment, professors felt a greater urgency to create new career opportunities for their students. Thus, they reached out to industry to set up joint projects: eventually, physicist L.S. Ornstein in Utrecht, for example, would deliver 42 Ph.D.'s to companies like Philips Electronics and the *Bataafsche Petroleum Maatschappij*, a subsidiary of Royal Dutch Shell.[30]

While a solid belief in the social benefits of science had already taken hold since the late nineteenth century, this had not immediately led to much research aimed directly at reaping these benefits.[31] However, the war had produced shortages of various kinds—like in the belligerent countries—and this was an impetus for the Dutch to find ways to make science help alleviate them. This effectively started up large-scale industrial research in Holland. The new emphasis was exhibited by the installation, in February of 1918, of the "Scientific Committee for Advice and Research in the interest of National Prosperity and Defence." It was headed by none other than Lorentz, and intended to fund the research reflected in its name.[32] Lorentz himself, like most of his generation and as a true theoretical physicist, had of course preferred, at his peak, to practice *science pour la science*, even though he also believed that its ultimate justification lies in the social and technological progress it could bring.[33] His committee was disbanded in 1922, as its results were judged as too poor and its focus as too academic. Instead, another committee headed by biologist F.A.F.C. Went concluded that the state needed to institute a central organization that more explicitly and effectively aimed directly at "applied science."[34]

---

[29] See Willink (1991) and Maas (2001) for explanations of this disproportionate success.
[30] Dorsman (2007), pp. 14-15; Heijmans (1994).
[31] Theunissen (1994).
[32] See e.g. Huijnen (2007), and references therein, and Huijnen's contribution to this volume. In 1918 Lorentz also became the chairman of another committee devoted to the closing off of the *Zuiderzee*, see Kox (2007).

Added to this shift in the desired role of science should be the complaints of scholars like Dijksterhuis and Kohnstamm, who observed an increase in anti-scientific and anti-intellectualist sentiments in Dutch culture in the 1920s.[35] Despite these pressures, most professors were still convinced that only free research and education, aiming at 'academic' values instead of practical careers, belonged at the universities. The proper place to do science was up in the ivory tower, from which *Bildung* would be handed down to students, and would eventually trickle down further into society.[36] Yet, how were such stances perceived beyond university campuses? Could they still be maintained and expect support in a society that was rapidly moving towards more democracy—universal suffrage for men was introduced in 1917—and increasingly demanded applicable knowledge? We will see in a short survey of Dutch newspapers and other non-academic publications whether these issues coloured the reception of relativity in any substantial way. Relativity, after all, seems an obvious example of elitist ivory tower science. The reception of relativity and Einstein may, in turn, also tell us more about the status of science in the post World War One years.

Important for the introduction of relativity to the Dutch public were the many public lectures held on the subject by Leiden's leading physicists between 1913 and 1919. These lectures would often find their way to the press and many were published in full. They usually took place before learned societies as the *Genootschap ter Bevordering van de Natuur-, Genees- en Heelkunde* (Society for the Advancement of Science, Medicine, and Surgery) in Amsterdam, or Rotterdam's *Bataafsch Genootschap der Proefondervinderlijke Wijsbegeerte* (Batavian Society for Experimental Philosophy). Audiences ranged from well-educated professionals to academic scholars.[37] Engineers were a particularly receptive audience and from their ranks rose several populaurisers and critics, such as Polak.[38]

Why did Lorentz, Ehrenfest and others put so much effort in popularising relativity theory? Apart from obvious reasons like the enjoyment it might give them, or a feeling of responsibility for sharing the most recent scientific insight with their fellow citizens, there was also a sense of urgency.[39] In 1922, Fokker, who had worked with Einstein in Zurich after having obtained his Ph.D. with Lorentz in 1913, wrote to the editors of *De Gids*: "popularization is an inescapable task in order to maintain the viability of science, when it is not aimed at applications."[40] In his opinion, fundamental physics in particular hung in the balance, which made it all the more necessary to get the "educated layperson [...] sympathetically and closely involved" in its discussions.[41] That such efforts were necessary but could have less of a result than was hoped for is exhibited by the report in the leftist-liberal magazine *De Groene Amsterdammer* about the sixteenth *Nederlandsch Natuur- en Geneeskundig Congres* (Dutch Scientific and Medical Congress) in 1917: the reporter applauded the recent effort to make the sciences more accessible to students, as this could counteract the highly inopportune shortages of qualified staff in industry. Having been greatly impressed by a lecture on war surgery, he further stated that medicine in particular constitutes "the most beautiful application of science" and that it was "more beautiful and useful than theories about the motion of heavenly bodies or the constitution of the atom."[42] Lorentz and Ehrenfest had been the star attractions at the congress: Lorentz had lectured on general relativity, and Ehrenfest had introduced the Bohr atom. Yet, the *Groene*'s reporter had apparently not become impressed.

---

[33] Theunissen & Klomp (1998).
[34] Van Rooij (2007).
[35] Klomp (1997), pp. 9-12; see Baneke (2008), especially pp. 39-42, for a slightly different point of view.
[36] Baneke (2006), pp. 32-36.
[37] Klomp (1997) lists a substantial number of these lectures, see p. 233; speakers included Lorentz, Ehrenfest and Fokker. Examples of their publications are: Lorentz (1913), (1915).
[38] See e.g. G.J. van de Well (1913); W.L. Brocades Zaalberg (1915); M.W. Polak (1918). Lorentz also published on relativity theory in *De Ingenieur*; see e.g. Lorentz (1917).

Economic pressures and the process of socio-cultural democratisation—we will return to this issue later when we discuss Klomp's account—led to a less than self-evident position for the sciences. Consequently, scientists saw a need to show themselves as accountable to society. By popularising the topic in lectures, theorists tried to justify their interest in the issues addressed by relativity. Relativity itself was in turn used, through its popularization, to legitimize the position of fundamental physics. Popularising was done in newspaper articles too, so we will consider these now.

## 4. Public reception

Public interest in relativity picked up greatly in the Netherlands after the 1919 confirmation of the gravitational bending of light, as it did elsewhere. The number of newspaper reports on Einstein and relativity quickly grew from a handful to over a hundred per year. Before 1919, relativity and Einstein were basically reported only in newspapers and magazines that were read by the academically educated elite, such as, in the case of newspapers, the *Algemeen Handelsblad*, the *Nieuwe Rotterdamsche Courant* or the *Nieuws van den Dag*.[43] Cultural journals like *De Gids* and *Onze Eeuw* regularly included articles on the sciences. Already before Eddington's announcement they had contained contributions on relativity by Lorentz and his Leiden colleague J.P. Kuenen.[44] After 6 November 1919, the date of Eddington's presentation, a broad spectrum of publications began including articles on relativity and particularly on Albert Einstein himself.[45] This may have been aided by Lorentz' praise for him in his newspaper article of 19 November, "Gravity and light. A confirmation of Einstein's theory of gravity."[46]

---

[39] Lorentz greatly enjoyed giving public lectures, according to his daughter (De Haas-Lorentz [1957], p. 86); Ehrenfest apparently complained about giving public lectures, yet he could use the extra income they provided (Hollestelle [2011], p. 189).
[40] Fokker to Kuenen, 31 July 1922; cited in Klomp (1997), pp. 54-55.
[41] Fokker in his book review of "Over den wereldaether" by J.D. van der Waals jr., cited in Klomp (1997), p. 54.
[42] R.T.A. Mees, "Zestiende Nederlandsch Natuur- en Geneeskundig Congres", *De Groene Amsterdammer*, 21 April 1917, p. 2.

By 1920 Einstein had become a celebrity in Holland, as he had elsewhere: his lectures, honours, troubles and job offers were eagerly reported, spurred on by the anti-relativity furor that had picked up in Germany that year.[47] That story was particularly closely covered, just like his appointment in Leiden and the associated travails.[48] Relativity's alleged incomprehensibility and its counterintuitive or avant-gardist nature very quickly became familiar tropes. For instance, in its report on the notorious French serial killer Henri Désiré Landru, *Het Vaderland* pointed out that "indeed, *rien ne se crée et rien ne se perd*, whatever Einstein and other Dadaists among the modern physicists may say: even light-hearted women do not evaporate completely into thin air if you burn them in the stove"[49] (that is, the material evidence in Landru's case had not successfully been made to disappear into nothing by the culprit). In any case, the idea was universally shared and was largely considered to be unproblematic among journalists that Einstein's general theory of relativity was, in the words of Lorentz, a "lasting monument of science".[50] Clearly, given that Einstein had become famous, relativity would be particularly suited for popularising and legitimising theoretical physics, while at the same time, these popularisations themselves would, of course, feed back into Einstein's fame.

---

[43] Dutch newspapers from the period can be accessed through the website of the *Koninklijke Bibliotheek* in The Hague; see http://kranten.kb.nl/ (accessed on 23 April 2013). For a description of their political position and readership, see Wijfjes (2004).
[44] See Lorentz (1915), Kuenen (1917).
[45] For example, titles as diverse as *Het Vaderland*, the local *Rotterdamsch Nieuwsblad* and the catholic paper *De Tijd* now regularly reported on Einstein.
[46] Lorentz (1919).
[47] See for example: "Einstein", 14 May 1920, *Het Vaderland* (announced Einstein's Leiden special professorship); "Einstein een Duitscher", 26 May 1920, *Algemeen Handelsblad* (informed readers of Einstein's birthplace and nationality); "Prof. Dr. Einstein", 27 August 1920, *Nieuwe Rotterdamsche Courant* (Einstein replied to his critics in the Berliner Tageblatt); "Prof. Einstein", 8 September, *Algemeen Handelsblad* (Einstein may receive the offer of a chair in Bern); "Een Einstein vergadering", 26 September 1920, *Algemeen Handelsblad* (on the debate with Philipp Lenard at Bad Nauheim; for more on the anti-relativity events in 1920, see the references in note 2).
[48] For example: "Prof. Einstein te Leiden", 18 May 1920, *Algemeen Handelsblad*; *Het Vaderland*; *De Tijd*; "prof. Einstein", 27 October 1920, *Leeuwarder Courant*; "Prof. Einstein", 18 October 1920, *De Sumatra post*.
[49] "Uit het laboratorium van Désiré Landru", 1 August 1920, *Het Vaderland*; for the incomprehensibility of relativity as a rhetorical element, see e.g. "De toestand", 24 November 1921, *Nieuwe Rotterdamsche Courant*.
[50] Lorentz (1919).

As discussed earlier, members of the Dutch elite were keen to portray themselves as internationalists and Einstein's role in Leiden could aid them in doing so, while they tried to realise their pacifist goals at the same time. Indeed, Einstein's participation in the 1920 conference on superconductivity in Kamerlingh Onnes' home, with Weiss and Langevin in attendance, was reported in a newspaper article with the headline: "Leiden as an international scientific centre."[51] The same piece also discussed a recent meeting, also in Leiden, between Einstein from Berlin, James Jeans from London, and a number of Dutch astronomers. Einstein himself was specifically identified with international reconciliation attempts: *Het Vaderland* approvingly reported on a lecture by Einstein at King's College, where its correspondent saw a "true attempt to bury a terrible past."[52] Einstein's trip to Paris in 1922 received coverage that was at least as extensive and again his reconciliation efforts were the main theme.[53] Clearly, the press was well aware that any Dutch attempts to achieve academic reconciliation would receive substantial support because of Einstein's public persona.

At the same time, Dutch reporters disapproved of the anti-relativity events in Germany, and saw the ills of Germany's old order and its rightist reactionaries with anti-Semitic motives reflected in those events.[54] In all, to reporters Einstein had very quickly become an iconic figure, who combined a rock solid reputation in science with progressive and internationalist politics, and a commitment to Jewish causes;[55] relativity was an achievement for which he should universally be revered and only ignorance or bigotry might obstruct such judgment. Dutch scientists under pressure would thus do well to show their professional and personal proximity to Einstein. By the end of 1920, after the news cycle on the British eclipse results, on Einstein's 'special' professorship in Leiden and on his reactionary opponents had run its course, Einstein had become a celebrity, an "idol".[56] A column in the Easter issue of *Het Vaderland* in April of 1922 related a "kaleidoscopic conversation" on modern culture between a lady and two gentlemen: when discussing sexologist Eugen Steinach, Oswald Spengler, theosophism, spiritism and transcendentalism, the conversation "came upon Einstein. How could you not talk about him?"[57]

---

[51] "Leiden als internationaal wetenschappelijk centrum", *Algemeen Handelsblad*, November 15, 1921.
[52] "Zomer in en buiten Londen", *Het Vaderland*, June 22, 1921.
[53] See for instance "Einstein over den internationale toestand", *Het Vaderland*, April 10, 1922.
[54] For example: "Studenten en politiek", *Algemeen Handelsblad*, 14 February 1920; "Heine-vereering", *Het nieuws van den dag voor Nederlandsch-Indië*, 20 November 1920; "Albert Einstein", *Het Vaderland*, 29 August 1920; "Broederschap", *Het Volk*, 3 September 1920: "Einstein", *Nieuw Israëlietisch Weekblad*, 3 September, 1920.
[55] On Einstein's Zionism, see e.g. "Opbouwfonds", *Algemeen Handelsblad*, 21 November 1919.
[56] As in *Wereldkroniek*, cited by *Nieuwe Rotterdamsche Courant*, 11 August 1921.
[57] "Kaleidoscoop", "Gesprek", *Het Vaderland*, 16 April 1922.

We have seen so far that, by and large, the Dutch public responded quite positively to Einstein in the years immediately after World War One. Science was under pressure to produce more applicable work, yet theoretical physics was still held in high regard, and in fact it could use relativity to exhibit a renewed relevance for itself that honoured Lorentz, while it also indicated novel intellectual horizons beyond his work. Einstein's close relationship with Leiden and its physicists of course aided this dynamic: it made it possible for Dutch audiences to identify with relativity even more strongly as it could be considered to be tied to local culture. At the same time, local scholarship extensively engaged with relativity and its creator, and shared this through the press. These circumstances are particular to the reception of relativity in the Netherlands.

Einstein's positions on the War were well known, and quite positively regarded. In fact, they resonated strongly with the Dutch, given their internationalist views. Einstein's negative judgment on the role of conservative German elites during the war was widely shared, and the anti-relativity actions were condemned equally as a misguided expression of "pan-Germanic" sentiments.[58] These circumstances are again particular to the Netherlands, but when considered from a broader perspective, they show once more that Einstein's, and subsequently relativity's reception was very much shaped by the highly politicized context of the war and its aftermath. In this sense, the Dutch case does resemble the reception histories of other European nations, even though the Netherlands had remained neutral, and its war years had been much less turbulent.

## 5. Relativity, democracy and educational reform

The Dutch reception of relativity was not universally positive, though. The theory soon engendered questions on the status of truth in scientific knowledge that played a key role in heated debates about the proper way to teach science in secondary schools. These debates were closely related to views on how society should ideally be organized: governed democratically, with the largest possible number of citizens enjoying a secondary education aimed at forming well rounded "persons", or ruled instead by a select and particularly competent group, steeped in abstract mathematically deduced truths, *à la* Plato. Those who held the latter position generally saw relativity as a threat. In what follows, we will outline the debate, basing ourselves on Henk A. Klomp's extremely valuable book *De relativiteitstheorie in Nederland. Breekijzer voor democratisering in het interbellum*[59], even if we do not necessarily agree with all its aspects.

---

[58] See "Heine-vereering", *Het nieuws van den dag voor Nederlandsch-Indië*, 20 November 1920.
[59] Klomp (1997).

Klomp's story has not been taken into consideration sufficiently outside the Netherlands, so a more extensive discussion of his work is warranted. At the same time, we would also like to include this account in a wider comparison, using what we have learnt from other reception studies.

In the debates that Klomp has described, the most interesting critic of relativity was philosopher-psychologist Gerard Heymans. Some of his positions were, in turn, of relevance to historian of science Eduard J. Dijksterhuis in his discussions on education. Dijksterhuis was employed as a mathematics teacher for most of his career, and his views on education were prompted first and foremost by his concern about mathematics education. Heymans, a Professor in Groningen, was a prominent intellectual of the ruling conservative-liberal elite. His views on knowledge and learning supported the social status quo before the introduction of universal suffrage in 1917. He believed that certain and objective judgments about nature could be made, and motivated this view by pointing to Kant's syntheses a priori and how these grounded mechanics; Newton's axioms and the constitutive role of Euclidean space, to Heymans, were definitive. Again inspired by Kant, Heymans further believed that, just as one could have certain knowledge of nature, one could also make moral judgments that were certain and objective, if one would only let the human ratio run its deductive course. Not everyone, however, could be expected to make such universal judgments: only those that had been steeped in the deductions of the sciences could be expected to see Plato's shadows and act morally. Only the intellectual elite, then, was suited to govern. Extending the right to vote would weaken the state by introducing non-objective judgments that were not aimed at the common good, which would make the state vulnerable to special interests. Heymans' conservatism was shared by many academics of his generation, such as influential historian and culture critic J. Huizinga, and to some extent also by Lorentz, although he was also a democrat.[60]

In the 1910s Heymans, already in his sixties, was confronted by scientists like Jacob Clay and Jan Schouten, who contended that his system was in conflict with the theory of relativity. Heymans reacted in 1921 by dismissing relativity in the pages of *De Gids*: Einstein had assumed that all knowledge was grounded in observation, thereby ignoring synthetic *a priori* judgments. The new facts that relativity could explain should be considered too small in number to "shake the foundations."[61] For Heymans, only explanations that gave real insight, based on obvious and evident foundations, could give the certainty of true knowledge. In essence, he simply compared relativity to his own epistemology, and found it wanting. Still, it must have been difficult for Heymans to argue against the new scientific consensus, given the authority that the sciences enjoyed in his system.

---

[60] On Lorentz and Heymans, see Theunissen & Klomp (1998); on Lorentz as a democrat, see e.g. Klomp (1997), p. 52.
[61] Heymans (1921), p. 98, as cited in Klomp (1997), p. 84.

Fokker responded immediately and strongly (he was something of a Dutch version of Hans Reichenbach[62]). He pointed out, for instance, that there was no need to consider absolute space a precondition to our knowledge of the world. Heymans' most relevant critic was physicist and pedagogue Philipp Kohnstamm. Kohnstamm maintained that Heymans illicitly imposed a human measure on nature when he insisted that physics needed to provide theories endowed with explanatory power. For Heymans, such theories gave certain knowledge, but Kohnstamm believed that one could possess no such thing: truth was like a person, whom one could meet, but never possess. He credited the insight that knowledge could not be certain to the appearance of the theory of relativity: natural laws had proven to be mere idealizations with only a limited validity. Such uncertainty was to be expected for knowledge grounded in experience and it also implied that one could not expect determinist certainty in moral judgments. This made room for religious and intuitive considerations in taking ethically just decisions (Kohnstamm, who was born Jewish, had been converted to Christianity in part because of his interactions with J.D. van der Waals sr.) Education should thus aim at forming emotionally rich and responsible characters, Kohnstamm argued in 1926: developing well rounded "personalities" was its essential task, according to his 'personalist' philosophy. Klomp points out that Kohnstamm placed a person's conscience above his ratio, so political power need not be restricted to those who have enjoyed a state education in accordance with Plato's ideals.[63] Kohnstamm was indeed a democrat: in fact, he had been party chairman of the *Vrijzinnig Democratische Bond* in 1917 that was partly responsible for introducing universal suffrage.

Fokker and Tatiana Ehrenfest-Afanassjewa, the wife of Paul Ehrenfest, presented ideas similar to Kohnstamm's. Both believed that intuitive and empirical reasoning should precede the introduction of abstract, deductive mathematics in education. They, too, pointed to relativity to justify their ideas. By the mid-1920s, Heymans had retired from criticizing relativity and his role was taken over by E.J. Dijksterhuis. Dijksterhuis was particularly concerned about the place of 'mechanics' in the Dutch secondary school curriculum. At the *Hogere Burgerschool* (the recently established advanced secondary school type that did not offer classical languages), mechanics had its own weekly four-hour slot, and was taught by mathematics teachers in a strictly deductive way, starting with Euclidean and Newtonian axioms. Physicists now appealed to the authority of relativity, claiming that mechanics was really an empirical science and that they should be the ones teaching it. The heated debate prompted the Ministry of Education to seek advice from, among others, Fokker and Dijksterhuis. In the end, it decided in 1934 to retain the mechanics course in its existing format.

---

[62] Reichenbach vigorously defended relativity against any criticism in the German language; see Hentschel (1990).
[63] Klomp (1997), p. 135-139.

Dijksterhuis' response to relativity was subtler than that of Heymans. He enlisted his historical scholarship in the hope of defusing its threat. Dijksterhuis argued that relativity was not nearly as innovative as had been claimed: since Galileo, the goal of physics had been to capture the simplest mathematical expression of the facts, and this process of mathematisation had brought the only true progress in the sciences. The process had reached its pinnacle with the formulation of Newton's axioms: Einstein's theory was just a recent addition.[64] Relativity still lacked a proper formal structure, and was, therefore, not suited to be presented in schools. Dijksterhuis strongly believed in Plato's educational philosophy, in which students were brought up in the strictest mathematical rigor: according to Dijksterhuis, deduction taught that claims needed to be substantiated, and built character. Empirical methods could not be a suitable replacement, and introducing relativity itself would undermine the deductive courses in mathematics, as it undermined their axioms. In 1937, Dijksterhuis ended up defending that the school curriculum should depart from the latest scientific consensus if that was necessary to retain Platonist ideals.

Dijksterhuis' influence was substantial in the 1920s due to, among other things, his membership in many advisory committees on educational matters, and it was boosted further when he became Secretary of the editorial board of *De Gids* in 1933. Kohnstamm criticized him for not caring about the dropouts of his selective mechanics courses, while for Dijksterhuis only educating the ruling elites seemed to matter. Dijksterhuis did feel intellectually at home with fascist ideologies and briefly joined the *Nationaal Front*—a marginal group that was fairly prominent in Dijksterhuis' home town of Oisterwijk—only to quickly resign his membership when he was confronted with the group's anti-Semitism.[65] In 1943, during the German occupation of the Netherlands, he accepted a professorship in history of science at the University of Amsterdam, which was considered a much graver offence after the occupation had ended. After 1945, Kohnstamm's ideas on education, and those of likeminded intellectuals such as Ehrenfest-Afanassjewa, gained much authority and in effect began to guide educational reform when the newly constituted Dutch Labour party (*Partij van de Arbeid*) embraced them in 1951.

It is interesting to note that Einstein himself, when expressing his pedagogical ideals, found, just like Kohnstamm, that schools should try first and foremost to form a "harmoniously developed person" with a "lively feeling for values," and that he pointed to the absence of authoritarian school systems in "democratically governed countries."[66] Kohnstamm emphasized

---

[64] Klomp believes that the debate on mechanics strongly influenced Dijksterhuis' arguments in his *Mechanization of the world picture* and *Val en worp* ('Free fall and projectile motion'); see Klomp, pp. 143, 183, 200. For Dijksterhuis' reaction to relativity, see also van Berkel (1996), p. 150.

[65] Van Berkel (1996), pp. 262-266.

[66] Albert Einstein, "Education for independent thought", *New York Times*, 5 October 1952, on p. 72 in Einstein (1994); and "On education", address at the tercentenary celebration of higher education in America, Albany, N.Y., 15 October 1936, pp. 63-69 in Einstein (1994), on p. 66. We thank Huub Rutjes for

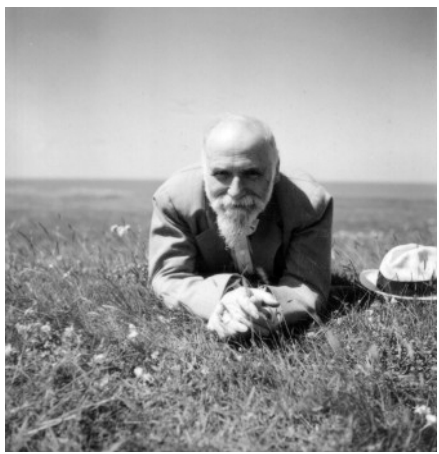
P.A. Kohnstamm in 1935, 60 years old, picture by his son, G.A. Kohnstamm. Source: Dolph Kohnstamm.

that drilling was the educational method of dictatorships, while Dijksterhuis held that hardly any student "is strong enough to be able to be free."[67] It is not too hard to imagine that a young Einstein might have been unhappy in Dijksterhuis' mechanics courses.

Thus, according to Henk Klomp, by inspiring Kohnstamm and others the theory of relativity eventually contributed to the democratisation of the Netherlands.[68] This conclusion is unproblematic, as far as we are concerned, when considered straightforwardly on the level of the historical facts. Yet, Klomp also raises the rhetorical question whether the course of events could have turned out different if it were not for the theory of relativity. His answer simply repeats the factual claim, thereby implicitly emphasizing the instrumental nature of the theory of relativity in this process of democratisation. Of course, elements of Einstein's theory went against certain presumptions of the epistemologies of Heymans and, though less so, of Dijksterhuis. Still, the contingencies in this story should not be overlooked. The Netherlands was already democratising due to forces far greater than the intellectual challenges posed by relativity. We can very well imagine scenarios in which other theories of physics would have played similar roles, as the determination to criticize conservative epistemologies was stronger than the force of arguments that might be mustered by relativity could ever be. Kohnstamm, for example, even claimed in 1926 that relativity introduced a new level of indeterminacy due to its inherently four-dimensional nature, despite personally communicated strong criticism of this point by Ehrenfest.[69] Clearly, Kohnstamm's aim was to dismiss epistemologies because they obstructed his pedagogy, rather than because he would like to convince Heymans, Dijksterhuis

---

drawing our attention to these articles. Also, one can easily see connections between Ehrenfest-Afanassjewa's ideas on education (Klomp 1997, pp. 166-169) and Einstein's epistemology, as for example found in the latter's essays "Geometry and Experience" (1921) and "On the method of theoretical physics" (1933), pp. 254-268 and pp. 296-303 in Einstein (1994).
[67] Klomp (1997), p. 208.
[68] Klomp (1997), p. 220.
[69] Klomp (1997), pp. 129-131.

or anyone else of the consequences of relativity theory. In all, the observation that the extensive public presence of relativity was partly due to inevitable philosophical debates that accompanied the Dutch democratisation process seems at least as justified as the notion that relativity intrinsically promoted that process and its debates.

Dijksterhuis, Heymans and the groups that they represented were not the only ones that objected to relativity. Klomp has reviewed its reception from the perspective of the social and ideological 'pillars' that typify Dutch society of the period.[70] Thus, he has further identified Catholic and Protestant critics, such as philosophers P.H.J. Hoenen and D.H.Th. Vollenhoven, and some socialist enthusiasts for relativity like Gerrit Mannoury. Similarly, our study of the newspapers of the period has taught us that responses in various groups were reflective of those groups' primary interests, though it should be added that Catholic or Protestant media did not report on Einstein or relativity critically, but simply less frequently or jubilantly. Addressing responses from the perspective of the ideological 'pillars' does not seem to add much beyond what one would expect, namely that these responses reflect Dutch 'pillarized' culture to some extent. Beyond the 'pillars', however, Klomp has also identified a common ground that critics of relativity shared: he agrees with the observations of philosopher and logician Evert Beth, who argued in 1964 that relativity had threatened the 'principle of obviousness', i.e. Aristotle's principle that the foundations of knowledge should be obvious, which played a central role in both Enlightenment and Christian philosophies. Thus, critics of relativity had defended their metaphysical beliefs by denying relativity proper authority in contradicting their epistemologies.[71]

Klomp has presented most of all a rich history of ideas, but he also draws attention to the fact that relativity's critics feared for the social positions and arrangements that were intellectually justified through their epistemologies.[72] This brings us to familiar territory: Dutch critics, though less radical, vehement, or explicitly political, were basically just the same as their anti-relativist counterparts in countries like Germany or the USA: they opposed the marginalization of their social and cultural values and positions. They believed that these had come under threat from the forces of modernization, forces whose other manifestations could range widely from abstract art or political change to specialization in the sciences, but which they identified directly or even only metaphorically with the theory of relativity.[73] Differences can, however, also be observed. As noted, the exchange of arguments in Holland was considerably more civilized than elsewhere, and there was virtually no vocal opposition from within the physics discipline itself that reached beyond the polite confines of academic debate. Circumstances that are particular to the Dutch context—the role of Lorentz and the ideals of neutrality and internationalism—have undoubtedly contributed to the moderate nature of the debate. In any case, it is most important to observe that Dutch opponents to relativity, like those who tried to obstruct changes in education or debated the necessity of *a priori* judgments, ultimately tried to resist social change and the downgrading of their positions and values. Their opposition thereby exhibited the same dynamic as that of anti-relativists in other post World War One societies.

---

[70] Klomp (1997), chapter 4, "Einstein en de verzuiling"; for an account of the 'process of pillarization' during the interwar years, see Lijphardt (1968).
[71] Klomp (1997), pp. 221-225.
[72] Klomp (1997), in particular p. 118.
[73] Recent literature that engages with this perspective is: Wazeck (2009), (2013); Van Dongen (2010), (2012). Interestingly, even Paul Ehrenfest, though a proponent of relativity and other theories at the vanguard of modernity, also struggled with some of its consequences; see van Lunteren & Hollestelle (2013).

## 6. Conclusions

The Dutch position in the international political context of the Great War and its aftermath played a substantial role in the presentation of Einstein and relativity to the public, and subsequently in how they were appreciated. The Dutch situation was quite different from that in neighbouring countries, yet it shares with them the circumstance *that* the reception of relativity was very much coloured by the war.

The prominent presence of relativity in Dutch society should be ascribed first of all to the culturally prominent position of the sciences in the period. Yet, as we have shown, that position came under pressure. In the case of relativity, this pressure actually aided the introduction of the theory further, as theoretical physicists like Fokker chose it to present their field; internationally prominent, yet closely tied to the Dutch tradition in physics, relativity seemed ideal for this purpose. Furthermore, Einstein's internationalist stances were appreciated by Dutch audiences, which again led to more, and more positive reporting on both Einstein himself and his theory. Finally, debates fuelled by the democratisation of Dutch society included relativity, as the theory was well suited for dismissing conservative epistemologies. Thus the public rise of relativity both resonated with, and accelerated Dutch society's democratising forces.

There was also criticism directed at relativity. Physicists who preferred the ether, like van der Waals jr., largely limited themselves to civilized if not predominantly academic debate— Van der Waals jr. actually expressed himself quite positively about relativity in the early 1920s[74]—, or, like Lorentz, they would still find many things to applaud Einstein and his theory for. Henk Klomp has shown that the most relevant public criticism originated in quarters that felt the theory was unjustly granted authority in epistemological matters. This sentiment was shared by a number of authors of various conservative persuasions, who, as we saw, resisted relativity, as they perceived it, as a threat to their values and the associated social positions or ideals. Although the Dutch opposition to relativity refrained from personal attacks, which was quite different elsewhere, we have argued that its underlying motivation derived from political and social frustrations or fears similar to those in other countries. Thus, the Dutch reception of relativity, though atypical in its particulars, was not atypical in an international comparison when viewed from the broader perspective of social change and its discontents. In sum, then, we conclude that the Netherlands has a reception history that departs from familiar stories of larger European nations in some respects , but resembles them in its underlying dynamics.

---

[74] Van der Waals jr. (1921), p. 84; (1923), p. 94. On van der Waals' positions in physics, see Maas (2001), pp. 151-155.

**References**

Baneke, D. (2006). Toegepaste natuurwetenschap aan de universiteit – contradictie of noodzaak? In L.J. Dorsman & P.J. Knegtmans (eds.), *Universitaire vormingsidealen: De Nederlandse universiteiten sedert 1876* (pp. 29-38). Hilversum: Verloren.

Baneke, D. (2008). *Synthetisch Denken: Natuurwetenschappers over hun rol in een moderne maatschappij, 1900-1940*. Hilversum: Verloren.

Biezunski, M. (1987). Einstein's reception in Paris in 1922. In T.F. Glick (ed.), *The comparative reception of relativity* (pp. 169-188). Dordrecht: Reidel Publishing Company.

Brocades Zaalberg, W.L. (1915). De cirkelgang van den aether. *Vragen van den Dag, 30*, 577-590.

CPAE 5. M.J. Klein, A.J. Kox & R. Schulmann, (eds.), *The Collected Papers of Albert Einstein*, Volume 5. *The Swiss years: Correspondence, 1902-1914*. Princeton: Princeton University Press (1993).

CPAE 8. R. Schulmann, A.J. Kox, M. Janssen, J. Illy & K. von Meyenn, (eds.), *The Collected Papers of Albert Einstein*, Volume 8. *The Berlin Years: Correspondence, 1914-1918.* Princeton: Princeton University Press (1998).

CPAE 9. D. Kormos Buchwald, R. Schulmann, J. Illy, D.J. Kennefick, T. Sauer, V.I. Holmes, A.J. Kox, & Z. Rosenkranz, (eds.), *The Collected Papers of Albert Einstein*, Volume 9. *The Berlin Years: Correspondence January 1919-April 1920*. Princeton: Princeton University Press (2004).

CPAE 10. D. Kormos Buchwald, T. Sauer, Z. Rosenkranz, J. Illy, V.I. Holmes, J. van Dongen, D.J. Kennefick, & A.J. Kox, (eds.), *The Collected Papers of Albert Einstein*, Volume 10. *The Berlin Years: Correspondence May-December 1920; Supplementary Correspondence 1909-1920*. Princeton: Princeton University Press (2006).

CPAE 13. D. Kormos Buchwald, J. Illy, Z. Rosenkranz, T. Sauer, J. van Dongen, D.J. Kennefick, A.J. Kox, D. Lehmkuhl, O. Moses, & I. Unna, (eds.), *The Collected Papers of Albert Einstein*,

Volume 13. *The Berlin Years: Writings and Correspondence January 1922-March 1923*. Princeton: Princeton University Press (2012).

Corduwener, P. (2012). Risee van de wereld of land van Grotius? De synthese tussen nationalisme en internationalisme in het Nederlandse fin de siècle. *Tijdschrift voor Geschiedenis, 125*, 202-215.

De Haas-Lorentz, G.L. (1957). Reminiscences. In De Haas-Lorentz, G.L. (ed.), *H.A. Lorentz, Impressions of his life and work* (pp. 82-120 ). Amsterdam: North-Holland.

De Sitter, W. (1911). On the bearing of the Principle of Relativity on Gravitational Astronomy. *Monthly Notices of the Royal Astronomical Society, 71*, 388-415.

Dorsman, L. (2007). Inleiding: Onderzoek in Opdracht. In L.J. Dorsman & P.J. Knegtmans, *Onderzoek in opdracht: De publieke functie van het universitaire onderzoek in Nederland sedert 1876* (pp. 9-21). Hilversum: Verloren.

Ehrenfest, P. (1912). *Zur Krise der Lichtäther-Hypothese*. Leiden: IJdo.

Einstein, A. (1920). *Äther und Relativitätstheorie. Rede gehalten am 5. Mai 1920 an der Reichs-Universität zu Leiden.* Berlin: Springer.

Einstein, A. (1994). *Ideas and opinions*. New York: The Modern Library.

Goenner, H. (1993). The Reaction to Relativity Theory I: The Anti-Einstein Campaign in Germany in 1920. *Science in Context*, *6*, 107-133.

Hentschel, K. (1990). *Interpretationen und Fehlinterpretationen der speziellen und der allgemeinen Relativitätstheorie durch Zeitgenossen Albert Einsteins*. Basel: Birkhäuser.

Heijmans, H.G. (1994). *Wetenschap tussen universiteit en industrie. De experimentele natuurkunde in Utrecht onder W.H. Julius en L.S. Ornstein 1896-1940*. Rotterdam: Erasmus.

Heymans, G. (1921). Leekenvragen ten opzichte van de relativiteitstheorie. *De Gids, 85 II*, 85-108.

Hollestelle, M. (2011). *Paul Ehrenfest: Worstelingen met de moderne wetenschap, 1912-1933.* Leiden: Leiden University Press.

Huijnen, P. (2007). Universiteit, bedrijfsleven en de opkomst van de beroepsonderzoeker 1880-1940. In L.J. Dorsman & P.J. Knegtmans (eds.), *Onderzoek in opdracht: De publieke functie van het universitaire onderzoek in Nederland sedert 1876* (pp. 23-37). Hilversum: Verloren.

Janssen, M. (2002). Reconsidering a scientific revolution: the case of Einstein versus Lorentz. *Physics in Perspective*, *4*, 421-446.

Kevles, D.J. (1971). 'Into hostile camps': The reorganization of international science in World War I. *Isis, 62*, 47-60.

Klein, M.J. (1970). *Paul Ehrenfest Volume 1: The making of a theoretical physicist*. Amsterdam: North-Holland.

Klomp, H.A. (1997). *De Relativiteitstheorie in Nederland: breekijzer voor democratisering in het interbellum*. Utrecht: Epsilon.

Kox, A.J. (1988). Hendrik Antoon Lorentz, the Ether, and the General Theory of Relativity. *Archive for History of Exact Sciences, 38*, 67-78.

Kox, A.J. (1992). General Relativity in the Netherlands, 1915-1920. In J. Eisenstaedt & A.J. Kox (eds.), *Studies in the History of General Relativity* (pp. 39-56). Boston: Birkhäuser.

Kox, A.J. (1993). Einstein and Lorentz. More than just good colleagues. *Science in Context, 6*, 43-56.

Kox, A.J. (2007). Uit de hand gelopen onderzoek in opdracht: H.A. Lorentz' werk in de Zuiderzee-commissie. In L.J. Dorsman & P.J. Knegtmans (eds.), *Onderzoek in opdracht: De publieke functie van het universitaire onderzoek in Nederland sedert 1876* (pp. 39-52). Hilversum: Verloren, 2007.

Kox, A.J. (ed.). (2008). *The Scientific Correspondence of H.A. Lorentz*. New York: Springer.

Kuenen, J.P. (1917). Relativiteits-theorie. *De Gids, 81 I*, 462-481; *81 II*, 96-123.

Lijphart, A. (1968). *The Politics of Accommodation: Pluralism and Democracy in the Netherlands*. Berkeley: University of California Press.

Lorentz, H.A. (1913). Nieuwe richtingen in de natuurkunde. *Nederlandsch Tijdschrift voor Geneeskunde, 57*, 2172-2183.

Lorentz, H.A. (1915). De lichtaether en het relativiteitsbeginsel: Voordracht, gehouden in de Verenigde Vergadering van de beide Afdeelingen der Akademie van Wetenschappen te Amsterdam, op 24 April 1915. *Onze Eeuw, 15 II*, 365-377.

Lorentz, H.A. (1917). De gravitatietheorie van Einstein en de grondbegrippen der natuurkunde. *De Ingenieur, 32*, 649-655.

Lorentz, H.A. (1919). De zwaartekracht en het licht. Een bevestiging van Einstein's gravitatietheorie. *Nieuwe Rotterdamsche Courant*, *19 November* (evening edition), 9-10.

Maas, A.J.P. (2001). *Individualisme en Atomisme: De Amsterdamse natuurkunde tussen 1877 en 1940*. Hilversum: Verloren.

Moatti, A. (2007). *Einstein, un siècle contre lui*. Paris: Odile Jacob.

Otterspeer, W. (1995). *Bolland. Een biografie.* Amsterdam: Bert Bakker.

Otterspeer, W. & J. Schuller tot Peursum-Meijer (1997). *Wetenschap en wereldvrede: De Koninklijke Akademie van Wetenschappen en het herstel van de internationale wetenschap tijdens het Interbellum*. Amsterdam: Koninklijke Nederlandse Akademie van Wetenschappen.

Paty, M. (1987). The Scientific Reception of Relativity in France. In T.F. Glick (ed.), *The comparative reception of relativity* (pp. 113-167). Dordrecht: Reidel Publishing Company.

Polak, M.W. (1918) *Bezwaren tegen de opvattingen der relativisten*. Deventer: Kluwer.

Rowe, D.E. (2006). Einstein's Allies and Enemies: Debating Relativity in Germany, 1916-1920. In V.F. Hendricks, K.F. Jørgensen, J. Lützen, & S.A. Pedersen (eds), *Interactions: Physics, Mathematics and Philosophy, 1860-1930* (pp. 231-280). Dordrecht: Springer.

Rowe, D.E. & Schulmann R. (eds.). (2007). *Einstein on politics. His private thoughts and public stands on nationalism, Zionism, war, peace and the bomb*. Princeton: Princeton University Press.

Sauer, T. (2007). Einstein and the early theory of superconductivity, 1919-1922. *Archive for History of Exact Sciences, 61*, 159-211.

Somsen, G. (2012). 'Holland's calling'. Dutch scientists' self-fashioning as international mediators. In R. Lettevall, G. Somsen & S. Widmalm (eds.), *Neutrality in twentieth century Europe. Intersections of science, culture, and politics after the First World War*. New York: Routledge, pp. 45-64.

Staley, R. (1998). On the Histories of Relativity: The Propagation and Elaboration of Relativity Theory in Participant Histories in Germany. *Isis, 89*, 263-299.

Stanley, M. (2003). 'An expedition to heal the wounds of war': The 1919 eclipse and Eddington as Quaker adventurer. *Isis, 94*, 57-89.

Tames, I. (2006). *Oorlog voor onze gedachten: Oorlog, neutraliteit en identiteit in het Nederlandse publieke debat 1914-1918*. Hilversum: Verloren.

Theunissen, B. (1994). Zuivere wetenschap en praktisch nut: visies op de maatschappelijk betekenis van wetenschappelijk onderzoek rond 1900. *Gewina, 17*, 141-144.

Theunissen, B. & Klomp H. (1998). H.A. Lorentz' visie op wetenschap. *Gewina, 21*, 1-14.

Van Berkel, K. (1996). *Dijksterhuis. Een biografie*. Amsterdam: Bert Bakker.

Van de Well, G.J. (1913). Het relativiteitsbeginsel in de mechanica. *De Ingenieur*, *38*, 800-802; 814-819; 839-845.

Van der Waals jr., J. D. (1909). *Over de vraag naar de meest fundamenteele wetten der natuur*. Groningen: Wolters.

Van der Waals jr., J. D. (1921). Over de ruimte. *Onze Eeuw*, *21 I*, 57-84.

Van der Waals jr., J. D. (1923). *De Relativiteits-Theorie*. Haarlem: De Erven F. Bohn.

Van Dongen, J. (2007). Reactionaries and Einstein's fame: 'German Scientists for the Preservation of Pure Science,' relativity, and the Bad Nauheim meeting. *Physics in Perspective, 9*, 212-230.

Van Dongen, J. (2010). On Einstein's opponents, and other crackpots. *Studies in History and Philosophy of Modern Physics*, *41*, 78-80.

Van Dongen, J. (2012). Mistaken Identity and Mirror Images: Albert and Carl Einstein, Leiden and Berlin, Relativity and Revolution. *Physics in Perspective, 14*, 126-177.

Van Kimmenade, A. (2010). *Resistance, resentment and relativity. A comparative analysis of French and German critics of the theory of relativity*. Master thesis, Utrecht University.

Van Lunteren, F. (2006). Wissenschaft internationalisieren: Hendrik Antoon Lorentz, Paul Ehrenfest und ihre Arbeit für die internationale Wissenschafts-Community. In A. Claussen (ed.), *Dokumentation. Einstein und Europa—Dimensionen moderner Forschung* (pp. 25-35). Düsseldorf: Wissenschaftszentrum Nordrhein-Westfalen.

Van Lunteren, F. & Hollestelle M. (2013). Paul Ehrenfest and the dilemmas of modernity. *Isis*, *104*, 504-536.

Van Rooij, A. (2007). Modellen van Onderzoek: De oprichting van TNO, 1920-1940. *Tijdschrift voor Sociale en Economische Geschiedenis*, *4*, 136-160.

Van Vollenhoven, C. (1910). Roeping van Holland. *De Gids*, *74 IV*, 185-204.

Voorhoeve, J.J.C. (1985). *Peace, profits and principles. A study of Dutch foreign policy*. Leiden: Martinus Nijhoff.

Warwick, A. (2003). *Masters of theory. Cambridge and the rise of mathematical physics*. Chicago: The University of Chicago Press.

Wazeck, M. (2009). *Einsteins Gegner: die öffentliche Kontroverse um die Relativitätstheorie in den 1920er Jahren*. Frankfurt a. M.: Campus.

Wazeck, M. (2013). Marginalization processes in science: The controversy about the theory of relativity in the 1920s. *Social Studies of Science*, *43*, 163-190.

Wijfjes, H. (2004). *Journalistiek in Nederland 1850-2000: beroep, cultuur en organisatie*. Amsterdam: Boom.

Willink, B. (1991). Origins of the Second Golden Age of Dutch Science after 1860: Intended and Unintended Consequences of Educational Reform. *Social Studies of Science, 21*, 503-526.